\documentclass[aps,prl,groupedaddress,twocolumn,draft,showpacs]{revtex4}
\usepackage{graphicx,epsf}
\newif\ifdraft
\drafttrue
\newif\ifpreprint
\preprinttrue

\def\fig#1{fig.~{\ref{#1}}}

\def\spa#1.#2{\left\langle#1\,#2\right\rangle}
\def\spb#1.#2{\left[#1\,#2\right]}
\def\tree{{\rm tree}}

\def\Tr{\, {\rm Tr}}
\def\SYM{MSYM}

\newbox\charbox
\newbox\slabox
\def\s#1{{      
        \setbox\charbox=\hbox{$#1$}
        \setbox\slabox=\hbox{$/$}
        \dimen\charbox=\ht\slabox
        \advance\dimen\charbox by -\dp\slabox
        \advance\dimen\charbox by -\ht\charbox
        \advance\dimen\charbox by \dp\charbox
        \divide\dimen\charbox by 2
        \raise-\dimen\charbox\hbox to \wd\charbox{\hss/\hss}
        \llap{$#1$}
}}

\def\eqn#1{eq.~(\ref{#1})}

\def\e{\epsilon}

\def\lr{\leftrightarrow}
\def\li#1{{\mathop{\rm Li}\nolimits}_#1}
\def\Li{\mathop{\rm Li}\nolimits}
\def\Split{\mathop{\rm Split}\nolimits}
\def\tree{{(0)}}
\def\Lloop{{(L)}}
\def\lloop{{(l)}}
\def\oneloop{{(1)}}
\def\twoloop{{(2)}}

\def\cg{\hat{c}_\Gamma}
\def\Ord{{\cal O}}

\begin{document}

\ifpreprint
hep-th/0309040
\hfill SLAC--PUB--10157
\hfill UCLA/03/TEP/23
\hfill Saclay-SPhT-T03/128
\fi

\title{Planar Amplitudes in Maximally Supersymmetric
Yang-Mills Theory}

\author{C. Anastasiou} 
\affiliation{{}Stanford Linear Accelerator Center, Stanford University, 
         Stanford, CA 94309} 

\author{Z. Bern} 
\affiliation{Department of Physics and Astronomy, UCLA, Los Angeles, CA
90095-1547} 

\author{L. Dixon} 
\affiliation{Stanford Linear Accelerator Center, Stanford University, 
         Stanford,  CA 94309} 

\author{D. A. Kosower} 
\affiliation{Service de Physique, Centre d'Etudes de Saclay, 
          F--91191 Gif-sur-Yvette cedex, France}

\date{September, 2003}

\begin{abstract}
The collinear factorization properties of two-loop scattering amplitudes in
dimensionally-regulated $N=4$ super-Yang-Mills theory suggest that, in
the planar ('t~Hooft) limit, higher-loop contributions can be
expressed entirely in terms of one-loop amplitudes.  We demonstrate
this relation explicitly for the two-loop four-point amplitude and,
based on the collinear limits, conjecture an analogous relation for
$n$-point amplitudes.  The simplicity of the relation is consistent
with intuition based on the AdS/CFT correspondence that the form of
the large-$N_c$ $L$-loop amplitudes should be simple enough to allow a
resummation to all orders.
\end{abstract}

\pacs{11.15.Bt, 11.25.Db, 11.25.Tq, 11.30.Pb, 11.55.Bq, 12.38.Bx
\hspace{1.0cm} } 

\maketitle



\renewcommand{\thefootnote}{\arabic{footnote}}
\setcounter{footnote}{0}

Four-dimensional quantum field theories are extremely intricate, and
generically have complicated perturbative expansions in addition to
non-perturbative contributions to physical quantities.  Gauge theories
are interesting in that numerous cancellations occur.  This renders
perturbative computations more tractable, and their results simpler,
than one might otherwise expect. The Maldacena
conjecture~\cite{Maldacena} implies that a special gauge theory is
simpler yet: the 't Hooft (planar) limit of maximally supersymmetric
four-dimensional gauge theory, or $N=4$ super-Yang-Mills theory (\SYM).
The conjecture states that the strong coupling limit of this conformal
field theory (CFT) is dual to weakly-coupled gravity in
five-dimensional anti-de Sitter (AdS) space.  The AdS/CFT
correspondence is remarkable in taking a seemingly intractable strong
coupling problem in gauge theory and relating it to a weakly-coupled
gravity theory, which can be evaluated perturbatively.  There have
been multiple quantitative tests of this correspondence, using
observables protected by supersymmetry (see {\it e.g.}
ref.~\cite{MaldacenaChecks}).  Because of the different domains of
validity of coupling expansions on the gauge and gravity sides,
quantitative comparisons involving unprotected quantities
rely at present on an additional expansion parameter, such as in the 
large-$J$ (``spin'') limit of BMN operators~\cite{BMN,MZBFST}.

In this latter context, the AdS/CFT correspondence can be used to motivate 
a search for patterns in the perturbative expansion of planar \SYM.
Intuitively, observables in the strongly coupled limit of this theory 
should be relatively simple because of the weakly-coupled gravity 
interpretation.  Yet infinite orders in the perturbative 
expansion, as well as non-perturbative effects, contribute to the strong
coupling limit.  How might such a complicated expansion organize itself
into a simple result?  For quantities protected by supersymmetry,
nonrenormalization theorems, or zeros in the perturbative series,
are one possibility.  Another possibility, for unprotected quantities,
is some iterative perturbative structure allowing for a resummation.
There have been some hints of an iterative structure developing in
the correlation functions of gauge-invariant composite
operators~\cite{Schubert}, but the exact structure, if it exists,
is not yet clear.

Amplitudes for scattering of on-shell (massless) quanta --- gluons,
gluinos, {\it etc.} --- are examples of particular interest because of
their importance in QCD applications to collider physics.  Although
the Maldacena conjecture does not directly refer to on-shell
amplitudes, we expect the basic intuition, that the perturbation
expansion should have a simple structure, to hold nonetheless.
Indeed, the simplicity of one- and two-loop amplitudes in \SYM\ has
allowed their computation to predate corresponding QCD
calculations~\cite{GSB,BRY}.

Perturbative amplitudes in four-dimensional massless gauge theories
are not finite, but contain infrared singularities due to soft and
collinear virtual momenta.  The divergences can be regulated using
dimensional regularization with $D=4-2\e$.  The resulting poles in
$\e$ begin at order $1/\e^{2L}$ for $L$ loops, and are described by
universal formul\ae{} valid for \SYM, QCD, {\it
etc.}~\cite{CataniDiv}.  To preserve supersymmetry we use the
four-dimensional helicity scheme~\cite{FDH} variant of dimensional
regularization, which is a close relative of dimensional
reduction~\cite{Siegel}.  The infrared divergences turn out to have
precisely the iterative structure we shall find in the full $N=4$
amplitudes; thus they provide useful guidance toward exhibiting such a
structure.

Infrared divergences generically prevent the definition of a
textbook $S$-matrix in a non-trivial conformal field theory such as \SYM.  
For the dimensionally regulated $S$-matrix 
elements we discuss, the regulator explicitly breaks the conformal 
invariance.  However, once the universal infrared singularities are 
subtracted, the four-dimensional limit of the remaining terms in the 
amplitudes may be taken, allowing an examination of possible connections 
to the Maldacena conjecture.  

These finite remainders are relevant for computing 
``infrared-safe'' observables in QCD, in which the divergent parts of 
virtual corrections cancel against real-radiative contributions 
(not discussed here) to produce finite perturbative 
results~\cite{LeeNauenberg}.  
The finite remainders should also be related to 
perturbative scattering matrix elements for appropriate coherent states
(see {\it e.g.} ref.~\cite{CEIR}).
The connection to the $S$-matrix for the true asymptotic states of
the theory, such as the hadrons of QCD, is of course non-trivial.

In \SYM, there are other hints that higher-loop amplitudes are related
in a simple way to the one-loop ones.  In particular, the integrands
of the amplitudes (prior to evaluation of loop-momentum integrals)
have a simple iterative structure~\cite{BRY}.  Furthermore, the
one-loop amplitudes have a relatively simple analytic structure, which
has allowed their determination to an arbitrary number of
external legs for configurations with maximal helicity
violation~\cite{Neq4} and up to six external legs for all
helicities~\cite{Fusing}.  Unitarity then suggests that higher-loop
amplitudes may also have a relatively simple analytic structure.

In this Letter we present direct evidence that this intuition is
correct for the planar amplitudes of \SYM.  A number of powerful techniques
are available to compute them.  These include the unitarity-based
method~\cite{Neq4,Fusing,BRY}; recently-developed multi-loop
integration methods (see ref.~\cite{MultiloopMethods} and references
therein); and the imposition of constraints from required behavior as
the momenta of two external legs become
collinear~\cite{CollinearLimitConstraints}.  Here we shall express the
explicit form for the four-point $N=4$ amplitude at two loops, in
terms of the one-loop amplitude, using previous
results~\cite{BRY,SmirnovDoubleBox}.  In addition, we present the
two-loop splitting amplitude in planar \SYM, computed elsewhere, which
summarizes the behavior of amplitudes as the momenta of two legs
become collinear.  We use the latter to provide evidence that the
relationship between the two-loop and one-loop amplitudes continues to
hold for an arbitrary number of external legs.

The leading-$N_c$ contributions to the $L$-loop $SU(N_c)$ gauge-theory
$n$-point amplitudes may be written as,
\begin{eqnarray}
{\cal A}_n^{(L)} & = & g^{n-2} 
 \biggl[ { 2 e^{- \e \gamma} g^2 N_c \over (4\pi)^{2-\e} } \biggr]^{L}
 \sum_{\rho} 
\Tr( T^{a_{\rho(1)}} 
   \ldots T^{a_{\rho(n)}} ) \nonumber \\
&& \null \hskip 1.3 cm \times
               A_n^{(L)}(\rho(1), \rho(2), \ldots, \rho(n))\,,
\end{eqnarray}
where the sum is over non-cyclic permutations of the external 
legs, and we have suppressed the momenta and helicities
$k_i$ and $\lambda_i$, leaving only the index $i$ as a label.
This decomposition holds for all particles in the gauge super-multiplet
because they are all in the adjoint representation.

\begin{figure}[t]
\centerline{\epsfxsize 3.2 truein \epsfbox{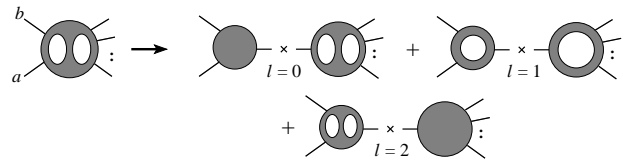}}
\caption[a]{\small The collinear factorization of a two-loop amplitude.
In each term in the sum, the left blob is a splitting amplitude,
and the right blob an $(n-1)$-point scattering amplitude.}
\label{FactFigure}
\end{figure}

The color-ordered amplitudes $A_n^{(L)}(1, 2, \ldots, n)$ satisfy 
simple properties as the momenta of two color-adjacent legs $k_a,k_b$
become collinear,
\begin{eqnarray}
 &&A_n^\Lloop(\ldots,a^{\lambda_a},b^{\lambda_b},\ldots)
  \, \longrightarrow \nonumber\\
&&\hphantom{A_n^\Lloop}\hskip -6mm
\sum_{l = 0}^L \sum_{\lambda=\pm}
  \Split^\lloop_{-\lambda}(z;a^{\lambda_a}\kern-1pt,b^{\lambda_b}) 
  A_{n-1}^{(L-l)}(\ldots,P^\lambda \kern-5pt,\ldots) \,. \hskip 1 cm 
\label{LoopSplit}
\end{eqnarray}
The index $l$ sums over the different loop orders of contributing
splitting amplitudes $\Split_{\lambda}^\lloop$, while $\lambda$ sums 
over the helicities of the fused leg $k_P=-(k_a + k_b)$, where $z$ is the
momentum fraction of $k_a$, $k_a = -z k_P$.  The two-loop version of
this formula is sketched in \fig{FactFigure}.  The splitting amplitudes 
are universal and gauge invariant.  Formula~(\ref{LoopSplit})
provides a strong constraint on amplitudes; for example, it has been used
to fix the form of a number of one-loop $n$-point
amplitudes~\cite{CollinearLimitConstraints,Neq4,Fusing}.

At tree level, the splitting amplitudes $\Split_{\lambda}^{(0)}$ are
the same in \SYM\ as in QCD.  Furthermore, the $N=4$ supersymmetry
Ward identities~\cite{SWI} imply that the \SYM\ loop splitting amplitudes
are all proportional to the tree-level ones, where the ratios depend
only on $z$ and $\e$, not on the helicity configuration, nor (except
for a trivial dimensional factor) on kinematic invariants~\cite{Neq4}.
We may therefore write the $L$-loop planar splitting amplitudes in
terms of ``renormalization'' factors $r_S^{(L)}(\e;z,s =
(k_1+k_2)^2)$, defined by
\begin{eqnarray}
 \Split_{-\lambda_P}^\Lloop(1^{\lambda_1},2^{\lambda_2})
& =&  \, 
  r_S^{(L)}
      \Split_{-\lambda_P}^\tree(1^{\lambda_1},2^{\lambda_2}) \,.
\label{LLooprDef}
\end{eqnarray}
Similarly defining the amplitude ratios 
$M_n^\Lloop(\e) \equiv A_n^\Lloop/A_n^\tree$,
we obtain in collinear limits,
\begin{eqnarray}
M_n^{\oneloop}(\e) &\rightarrow& M_{n-1}^{\oneloop}(\e) + r_S^{(1)}(\e) \,,
\label{OneLoopCollinear}\\
M_n^{\twoloop}(\e) & \rightarrow & M_{n-1}^{\twoloop}(\e)
+ r_S^{(1)}(\e) M_{n-1}^{\oneloop}(\e) + r_S^{(2)}(\e) \,. \hskip .5 cm 
\label{TwoloopCollinear}
\end{eqnarray}

The $N=4$ one-loop splitting amplitudes have been
calculated to all orders in $\e$~\cite{OneLoopSplitting}, with the result
\begin{eqnarray}
r_S^{(1)}(\e;z,s) &=&  { \cg  \over \e^2} 
  \biggl( { \mu^2 \over -s } \biggr)^\e
  \biggl[ - {\pi \e \over \sin(\pi \e) } \biggl( { 1-z \over z } \biggr)^\e
\nonumber \\
&& \hskip3mm \null
          + 2 \, \sum_{k=0}^\infty \e^{2k+1} \,
                   \Li_{2k+1}\biggl( {- z \over 1-z } \biggr) \biggr] \,,
\label{OneLooprSUSY}
\end{eqnarray}
where $\Li_n$ is the $n$-th polylogarithm, 
\begin{equation}
\cg = {e^{\e \gamma} \over 2} 
{\Gamma(1+\e) \Gamma^2(1-\e) \over \Gamma(1-2\e) } \,,
\label{rGammaDef}
\end{equation}
and $\gamma$ is Euler's constant.

We have calculated the two-loop, leading-$N_c$, $N=4$ splitting amplitudes 
through $\Ord(\e^0)$ using the method of ref.~\cite{KosowerSplit}
with the result,
\begin{equation}
r^{(2)}_S(\e;z, s)  =   {1 \over 2} \bigl( r^{(1)}_S(\e;z, s) \bigr)^2
     + f(\e) r^{(1)}_S(2\e;z, s) \,,
\label{OneloopTwoloopSplit}
\end{equation}
where 
\begin{equation}
f(\e) \equiv (\psi(1-\e)-\psi(1))/\e
= - (\zeta_2 + \zeta_3 \e + \zeta_4 \e^2 + \cdots)
\label{fDef}
\end{equation}
with $\psi(x) = (d/dx) \ln\Gamma(x)$, $\psi(1) = -\gamma$.

The infrared singularities of leading-$N_c$ \SYM\
at one and two loops can be extracted from more general studies, notably 
ref.~\cite{CataniDiv}.  At one loop, the divergences are given by,
\begin{equation}
C_n^\oneloop(\e) = - {e^{\e \gamma} \over 2 \Gamma(1-\e)} {1\over
    \e^2} \sum_{i=1}^n \biggl( {\mu^2 \over -2 k_i \cdot k_{i+1}}
    \biggr)^\e \,.
\label{OneloopIRDivergence}
\end{equation}
The two-loop divergences, in the four-dimensional
helicity scheme, are~\cite{CataniDiv,Twoloopgggg},
\begin{eqnarray}
C_n^\twoloop(\e) &= & {1\over 2} \Bigl(C_n^\oneloop(\e)\Bigr)^2 + 
                C_n^\oneloop(\e) F_n^\oneloop(\e)  \nonumber \\
&& \hskip-2mm
           - ( \zeta_2 + \e \zeta_3 ) 
              { e^{-\e \gamma} \Gamma(1-2\e) \over \Gamma(1-\e) }
                   C_n^\oneloop(2\e).
\label{TwoloopIRDivergence}
\end{eqnarray}
The finite remainder is defined by subtraction,
\begin{equation}
 F_n^\Lloop(\e) = M_n^{\Lloop}(\e) - C_n^\Lloop(\e) \,.
\label{FiniteLloop}
\end{equation}
Note that $C_n^\Lloop(\e)$ contains some finite terms as well.


We now present evidence that through $\Ord(\e^0)$
the two-loop planar amplitudes are 
related to one-loop ones via,
\begin{equation}
M_n^{\twoloop}(\e) =  {1 \over 2} \Bigl(M_n^{\oneloop}(\e) \Bigr)^2
             + f(\e) \, M_n^{\oneloop}(2\e) - {5\over 4} \zeta_4 \,.
\label{TwoloopOneloop}
\end{equation}
Note the similarity of our ansatz to the two-loop splitting
amplitude~(\ref{OneloopTwoloopSplit}), as well as to the infrared
subtraction (\ref{TwoloopIRDivergence}). 

\begin{figure}[t]
\centerline{\epsfxsize 1.8 truein \epsfbox{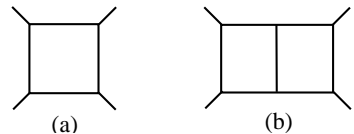}}

\vskip -.2 cm 
\caption[a]{\small The scalar integral functions appearing
in the (a) one- and (b) two-loop four-point $N=4$ amplitudes.}
\label{IntegralFigure}
\end{figure}

The one-loop four-point amplitude in \SYM\
was first calculated by taking the low energy limit
of a superstring~\cite{GSB}.  The result is given in terms of a
one-loop scalar box diagram, 
depicted in \fig{IntegralFigure}(a).
(This integral is identical to the one appearing in scalar
$\phi^3$ theory.)  
Expanding the result in $\e$ yields, 
\begin{eqnarray}
M_4^{\oneloop}(\e) \!\! &= &  \!\!\cg \biggl\{
- {2\over \e^2} \biggl({\mu^2 \over -s} \biggr)^{\e}  
- {2\over \e^2} \biggl({\mu^2 \over -t} \biggr)^{\e}\nonumber \\
&& \null 
+ \biggl({\mu^2 \over u} \biggr)^{\e} 
      \biggl[ 
   {1\over2} \Bigl( (X-Y)^2+ \pi^2 \Bigr)
\nonumber\\ && \null 
 + 2 \e  \biggl( \li3(x) - X \li2(x) 
              - {X^3 \over 3} 
          - {\pi^2\over2} X \biggr)
\nonumber\\ &&  \null 
 - 2 \e^2 \biggl( \li4(x) + Y  \li3(x)
             - {X^2 \over2} \li2(x)
             - {X^4\over8}  
\nonumber\\ && \null 
             - {X^3 Y \over6}  
            + {X^2 Y^2 \over4} 
            - {\pi^2\over4} X^2  - {\pi^2\over3} X Y  
              - 2 \zeta_4 \biggr) 
\nonumber\\ && \null 
 \ +\ \hbox{$(s \lr t)$}\ \biggr]\ +\ \Ord(\e^3) \biggr\} \, , 
\end{eqnarray}
where $s=(k_1+k_2)^2$,  $t=(k_1+k_4)^2$, $u=-s-t$, $x = -s/u$, 
$y=-t/u$, $X = \ln x$, and $Y = \ln y$.
For the four-point case, the $\e \rightarrow 0$ limit of the 
finite remainder~(\ref{FiniteLloop}) is
\begin{equation}
F_4^\oneloop(0) =
 {1\over 2} 
\ln^2 \biggl( {-s\over -t} \biggr) + {\pi^2 \over 2} \,.
\label{OneLoopFinite}
\end{equation}

In ref.~\cite{BRY} the two-loop $N=4$ amplitude was presented in terms
of a double-box scalar integral depicted in \fig{IntegralFigure}(b),
plus its image under the permutation $s \lr t$.
Ref.~\cite{SmirnovDoubleBox} provides the explicit value of this
integral, through $\Ord(\e^0)$, in terms of polylogarithms.  Inserting
this value, we obtain precisely the result (\ref{TwoloopOneloop}) with
$n=4$.  The equality requires the use of polylogarithmic identities,
and involves a non-trivial cancellation of terms between the two
contributing integrals.  Terms through $\Ord(\e^2)$ in
$M_4^{\oneloop}(s,t)$ contribute at $\Ord(\e^0)$ in
$M_4^{\twoloop}(s,t)$, since they can multiply the $1/\e^2$ terms.

Subtracting the two-loop infrared divergence given
in \eqn{TwoloopIRDivergence} from our calculated expression
yields 
\begin{equation}
F_4^\twoloop(0) = 
{1\over 2} \Bigl[ F_4^\oneloop(0) \Bigr]^2 - \zeta_2 \, F_4^\oneloop(0) 
            - {21\over8} \zeta_4 \,,
\label{TwoloopOneloopFinite}
\end{equation}
expressed in terms of the one-loop finite remainder (\ref{OneLoopFinite}).

For $n\ge 5$ legs, we examine the properties as external momenta
become collinear, using eq.~(\ref{LoopSplit}).  Applying the one-loop
collinear behavior (\ref{OneLoopCollinear}) to the ansatz
(\ref{TwoloopOneloop}), we have
\begin{eqnarray}
M_n^\twoloop(\e) &\rightarrow&
  {1\over 2} \Bigl(M_{n-1}^\oneloop(\e) + r_S^{(1)}(\e)\Bigr)^2 
\nonumber\\
&&\hphantom{ {1\over2} \hskip-2mm}
      + f(\e) \bigl[M_{n-1}^\oneloop(2\e) + r_S^{(1)}(2\e) \bigr] 
    - {5 \over 4} \,  \zeta_4\,, \hskip 1 cm 
\end{eqnarray}
which is consistent with the required two-loop collinear
properties~(\ref{TwoloopCollinear}), using \eqn{OneloopTwoloopSplit}.
Although severely constrained, amplitudes are not uniquely defined by
their collinear limits~\cite{Neq4}.  Thus \eqn{TwoloopOneloop} remains
unproven for $n\geq5$.  The direct computation of the two-loop
five-point function seems feasible, and would provide an important
test of the ansatz.

We investigated two potential extensions of the 
relation~(\ref{TwoloopOneloop}), each with negative results:
\par\noindent
1) We examined the non-planar extension by computing the subleading-color
two-loop finite remainders, analogous to $F_4^{(2)}(0)$.  These terms contain
polylogarithms, and hence cannot be written in terms of one-loop
finite remainders, unlike the planar \eqn{TwoloopOneloopFinite}.
Thus the non-planar terms do not appear to have a structure analogous to
\eqn{TwoloopOneloop}, in line with heuristic expectations from the Maldacena 
conjecture.
\par\noindent
2) For the four-point amplitude, we find that \eqn{TwoloopOneloop} is {\it not}
satisfied at $\Ord(\e)$, due to polylogarithmic obstructions.
Hence the relation holds only as $D \rightarrow 4$, {\it i.e.} where
the theory becomes conformal.

The possibility of resumming perturbative expansions in \SYM\ may also
have relevance for QCD.  QCD may be viewed as containing a ``conformal
limit'' ({\it e.g.}  \SYM) plus conformal-breaking terms.  This
perspective has had practical impact on topics ranging from the
Crewther relation to exclusive processes~\cite{ConformalLimitQCD}.  We
remark that $N=4$ amplitudes can be obtained directly from QCD
amplitudes by adjusting the number and color of states circulating in
the loop: starting from the two-loop QCD amplitudes of
ref.~\cite{Twoloopgggg} and substituting for the `spin index dimension' $D_s
= 4 - 2\e \delta_R \rightarrow 10$ and for the color Casimirs $C_F \rightarrow
C_A, T_R N_f \rightarrow 2 C_A$, one obtains the two-loop \SYM\
amplitudes.  (These modifications effectively give $D=10$, $N=1$
super-Yang-Mills theory, truncated to $D=4$, which is $N=4$
super-Yang-Mills theory.  See eq.~(6.5) of ref.~\cite{Twoloopgggg} for
an analogous conversion to $D=4$, $N=1$ super-Yang-Mills amplitudes.)

A number of open questions deserve further study.  At two loops, the
$N=4$ planar ansatz should be checked for at least the five-point
case.  At higher loop order, the intuition described in the
introduction suggests a continuation of the iterative structure found
at two loops, possibly enabling a resummation of perturbative
contributions.  Thus we expect that higher-loop planar $N=4$
amplitudes will be ``polynomial'' functions of the one-loop
amplitudes.  Indeed, the known form of three-loop infrared
divergences~\cite{CataniDiv} provides some confirmation of this.
Recent advances should make possible explicit evaluation of the
leading-color three-loop four-point amplitude, using known expressions
for the integrand~\cite{BRY}.  (One of the two three-loop integrals
needed has already been computed through $\Ord(\e^0)$ in terms of
generalized polylogarithms~\cite{SmirnovTripleBox}.)  One would also
like to identify a symmetry (presumably related to super-conformal
invariance) and associated Ward identity responsible for restricting
amplitudes to be iterations of the one-loop amplitude; recall that the
relation between the two-loop and one-loop amplitudes holds only near
$D=4$ where the theory is conformal.  We are optimistic that an
understanding of the amplitudes of $N=4$ super-Yang-Mills theory will
lead to new insight into consequences of the AdS/CFT correspondence.

We thank E.~D'Hoker, P.~Kraus, and E.T.~Tomboulis for helpful
discussions.  This research was supported by the US Department of
Energy under contracts DE-FG03-91ER40662 and DE-AC03-76SF00515 and by
the {\it Direction des Sciences de la Mati\`ere\/}
of the {\it Commissariat \`a l'Energie Atomique\/} of France.



\vskip -.4  cm
\begin{thebibliography}{99}

\bibitem{Maldacena}
J.M.~Maldacena,
Adv.\ Theor.\ Math.\ Phys.\ {\bf 2}, 231 (1998)
[Int.\ J.\ Theor.\ Phys.\ {\bf 38}, 1113 (1999)]
[hep-th/9711200];
%
S.S.~Gubser, I.R.~Klebanov and A.M.~Polyakov,
Phys.\ Lett.\ B {\bf 428}, 105 (1998)
[hep-th/9802109];
%
O.~Aharony, S.S.~Gubser, J.M.~Maldacena, H.~Ooguri and Y.~Oz,
Phys.\ Rept.\  {\bf 323}, 183 (2000)
[hep-th/9905111].

\bibitem{MaldacenaChecks}
E.~D'Hoker and D.Z.~Freedman,
hep-th/0201253.

\bibitem{BMN}
D.~Berenstein, J.~M.~Maldacena and H.~Nastase,
JHEP {\bf 0204}, 013 (2002).

\bibitem{MZBFST} 
J.A.~Minahan and K.~Zarembo,
JHEP {\bf 0303}, 013 (2003)
[hep-th/0212208];
N.~Beisert, S.~Frolov, M.~Staudacher and A.A.~Tseytlin,
JHEP {\bf 0310}, 037 (2003),
and references therein.

\bibitem{Schubert}
B.~Eden, P.S.~Howe, C.~Schubert, E.~Sokatchev and P.C.~West,
Phys.\ Lett.\ B {\bf 466}, 20 (1999)
[hep-th/9906051];
%
B.~Eden, C.~Schubert and E.~Sokatchev,
Phys.\ Lett.\ B {\bf 482}, 309 (2000)
[hep-th/0003096];
hep-th/0010005.

\bibitem{GSB}
M.B.~Green, J.H.~Schwarz and L.~Brink,
Nucl.\ Phys.\ B {\bf 198}, 474 (1982).

\bibitem{BRY}
Z.~Bern, J.S.~Rozowsky and B.~Yan,
Phys.\ Lett.\ B {\bf 401}, 273 (1997)
[hep-ph/9702424];
%
Z.~Bern, L.J.~Dixon, D.C.~Dunbar, M.~Perelstein and J.S.~Rozowsky,
Nucl.\ Phys.\ B {\bf 530}, 401 (1998)
[hep-th/9802162].

\bibitem{CataniDiv}
S.~Catani,
Phys.\ Lett.\ B {\bf 427}, 161 (1998) [hep-ph/9802439];
%
G.~Sterman and M.E.~Tejeda-Yeomans,
Phys.\ Lett.\ B {\bf 552}, 48 (2003)
[hep-ph/0210130].

\bibitem{FDH}
Z.~Bern and D.A.~Kosower,
Nucl.\ Phys.\ B {\bf 379}, 451 (1992);
Z.~Bern, A.~De Freitas, L.~Dixon and H.L.~Wong,
Phys.\ Rev.\ D {\bf 66}, 085002 (2002) [hep-ph/0202271].

\bibitem{Siegel}
W.~Siegel,
Phys.\ Lett.\ B {\bf 84}, 193 (1979).

\bibitem{LeeNauenberg}
T.~Kinoshita, 
J.\ Math.\ Phys. {\bf 3}, 650 (1962);
T.D.~Lee and M.~Nauenberg,
Phys. Rev. {\bf 133}, B1549 (1964).

\bibitem{CEIR}
H.F.~Contopanagos and M.B.~Einhorn,
Phys.\ Rev.\ D {\bf 45}, 1291 (1992).

\bibitem{Neq4}
Z.~Bern, L.J.~Dixon, D.C.~Dunbar and D.A.~Kosower,
Nucl.\ Phys.\ B {\bf 425}, 217 (1994)
[hep-ph/9403226].

\bibitem{Fusing}
Z.~Bern, L.J.~Dixon, D.C.~Dunbar and D.A.~Kosower,
Nucl.\ Phys.\ B {\bf 435}, 59 (1995)
[hep-ph/9409265].

\bibitem{MultiloopMethods}
T.~Gehrmann and E.~Remiddi,
in {\it Proc. of the 5th International Symposium on 
Radiative Corrections (RADCOR 2000) }, ed. Howard E. Haber,
hep-ph/0101147.

\bibitem{CollinearLimitConstraints}
Z.~Bern, G.~Chalmers, L.J.~Dixon and D.A.~Kosower,
Phys.\ Rev.\ Lett.\  {\bf 72}, 2134 (1994)
[hep-ph/9312333].

\bibitem{SmirnovDoubleBox}
V.A.~Smirnov,
Phys.\ Lett.\ B {\bf 460}, 397 (1999)
[hep-ph/9905323].

\bibitem{SWI}
M.T.~Grisaru, H.N.~Pendleton and P.~van Nieuwenhuizen,
Phys.\ Rev.\ D {\bf 15}, 996 (1977);
M.T.~Grisaru and H.N.~Pendleton,
Nucl.\ Phys.\ B {\bf 124}, 81 (1977).

\bibitem{OneLoopSplitting}
Z.~Bern, V.~Del Duca and C.R.~Schmidt,
Phys.\ Lett.\ B {\bf 445}, 168 (1998)
[hep-ph/9810409];
%
D.A.~Kosower and P.~Uwer,
Nucl.\ Phys.\ B {\bf 563}, 477 (1999)
[hep-ph/9903515];
%
Z.~Bern, V.~Del Duca, W.B.~Kilgore and C.R.~Schmidt,
Phys.\ Rev.\ D {\bf 60}, 116001 (1999)
[hep-ph/9903516].

\bibitem{KosowerSplit}
D.A.~Kosower,
Nucl.\ Phys.\ B {\bf 552}, 319 (1999)
[hep-ph/9901201].

\bibitem{Twoloopgggg}
Z.~Bern, A.~De Freitas and L.~Dixon,
JHEP {\bf 0203}, 018 (2002)
[hep-ph/0201161].

\bibitem{ConformalLimitQCD}
R.J.~Crewther,
Phys.\ Rev.\ Lett.\  {\bf 28}, 1421 (1972);
S.J.~Brodsky, E.~Gardi, G.~Grunberg and J.~Rathsman,
Phys.\ Rev.\ D {\bf 63}, 094017 (2001)
[hep-ph/0002065];
B.~Meli\'c, D.~M\"uller and K.~Passek-Kumeri\v{c}ki,
Phys.\ Rev.\ D {\bf 68}, 014013 (2003)
[hep-ph/0212346].

\bibitem{SmirnovTripleBox}
V.A.~Smirnov,
Phys.\ Lett.\ B {\bf 567}, 193 (2003)
[hep-ph/0305142].

\end{thebibliography}
\end{document}